# Prisec II - A Comprehensive Model for IoT Security: Cryptographic Algorithms and Cloud Integration

P.M.C.F da Costa , V. R. Q. Leithardt (Senior Member, IEEE)

*Abstract—* This study addresses the critical issue of ensuring data security and efficiency in interconnected devices, especially in IoT environments. The objective is to design and implement a model using cryptographic algorithms to enhance data security in 5G networks. Challenges arise from the limited computational capabilities of IoT devices, which require the analysis and selection of cryptographic algorithms to achieve efficient data transmission. This study proposes a model that includes four levels of security, each employing different levels of encryption to provide better data security. Finally, cloud computing is used to optimize processing efficiency and resource utilization to improve the overall data transmission.

## I. Introduction

With the increasing prevalence of cloud computing, Edge and Big Data in both everyday life and business operations, ensuring security of data processed by IoT devices against potential intruders has become a paramount concern. The fields of medicine, smart grids, home automation, precision agriculture and urban mobility represent just a few of the many scenarios where cloud computing is present [5]. However, with the benefits of connectivity also come significant risks, particularly regarding data security, which is the focus of this study.

In this context, cryptography plays a vital role in safeguarding sensitive information since this technique consists of developing and using coded algorithms to protect and obscure transmitted information so that it may only be read by those with the permission and ability to decrypt it [16]. In summary, cryptography plays a significant role in data security ensuring that the data transmitted amidst interconnected system stay protected.

Cloud computing, on the other hand, refers to the delivery of computing services including storage, databases, networking, and other resources over the internet. It offers numerous benefits, such as security, speed, cost, and others. There are numerous uses for cloud computing such as test and build applications, stream audio and video, analyze data and others [15]. In this study, the cloud will be used for testing and build application together with some cryptography algorithms, being this explained further in this paper.

Nowadays any IoT application can be divided into four layers: sensing layer, network layer, middleware layer and application layer. Each of these layers in an IoT application uses diverse technologies that bring a number of issues and security threats. The application layer directly deals with and provides services to end users and in this layer, there are security issues that are not present in the other layers, such as data theft and DDoS attacks. IoT applications deal with a lot of critical and private data and there is a lot of data movement, with that being the case the data is even more vulnerable to attacks than data at rest, one of the methods to secure IoT applications against this threat is data encryption [15]. With that being said, this study endeavors to address these concerns of data security through the implementation of cryptography, in achieve that it's necessary to search for cryptographic algorithms that are used nowadays, especially in scenarios and applications that use 5G networks, and test them considering factors such as packet quantity and size. By focusing on the development and deployment of a robust cryptographic solution, the study aims to increase the security of IoT devices and protect sensitive data from unauthorized access and tampering.

The first goal is to select and analyze different cryptographic algorithms to identify the algorithms with the best performance in terms of encryption and decryption time, while considering the size of the packet. The next objective is to implement these algorithms in the structure model that will be further explained in the report to evaluate the time taken to encrypt and decrypt different sizes of packets. After finding the best structure model, the next objective is to analyze the performance of this model regarding the number of packets transmitted, after that this structure model will be used together with cloud computing while also evaluating the performance of the model.

## II. Related works

It is common knowledge that, security is an increasing concern in real-time systems, so researchers have demonstrated attacks and defenses aimed at such systems. In the article [12], it was identified, classified and measured the effectiveness of the security research in this domain. It was provided a high-level summary [identification] and a taxonomy [classification] of this existing body of work. Furthermore, it was carried out an in-depth analysis [measurement] of scheduler-based security techniques — the most common class of real-time security mechanisms.

For this purpose, it was developed a common metric, "attacker's burden", used to measure the effectiveness scheduler-based real-time security measures.

Also in the security topic, it is important to talk about decentralized systems that are systems that disperse computation tasks to multiple parties without relying on a trusted central authority. Since any party can be compromised by attackers, ensuring security in these systems is a crucial task therefore in the article [6], it was examined prior security solutions and study the inherent difficulties of securely performing computation tasks in decentralized systems by focusing on three complementary components. It was evaluated the performance of cryptographic algorithms in decentralized systems where nodes may have different amounts of computing resources, while also providing a benchmark of widely deployed cryptographic algorithms on devices with a different extent of resource constraints and show what computing capabilities are required for a device to perform expensive cryptographic

As stated in [6] and [12], security is very important especially when dealing with data and according to the article [1], with the proliferation of networked systems and our increasing reliance on technology, the risk of data breaches and unauthorized access has escalated. Therefore, the development of robust cryptographic frameworks is imperative. To confront this challenge, this article proposes a cryptographic model that integrates encryption, hash functions, and digital signatures to safeguard data from unauthorized access, ensure data integrity, and establish secure communication channels. Following this line of thought, numerous papers have developed cryptographic frameworks to address security vulnerabilities. Firstly, for the security vulnerabilities in wireless sensor networks (WSNs), it was introduced in [4] an advanced cryptographic framework designed to tackle these vulnerabilities head-on. The heart of this framework lies in the seamless integration of two potent cryptographic forces: Elliptic Curve Cryptography (ECC) and the Advanced Encryption Standard (AES). The result is the ECC and AES cryptographic amalgamation, positioned as a robust solution to fortify the security of data transmission within WSNs. In [13] is presented the evaluation of symmetric key ciphers, including Advanced Encryption Standard (AES), Rivest Cipher 6 (RC6), Twofish, SPECK128, LEA, and ChaCha20-Poly1305.

Through rigorous experimentation, researchers have scrutinized execution times, throughput, and power consumption to discern optimal algorithms for securing IoT communications. Furthermore, the impact of cryptographic algorithms on low-computational devices commonly found in smart home environments has been thoroughly examined in [14]. Studies done in six commonly used embedded devices in IoT WSNs: ESP8622, ESP32, and Raspberry Pi (RPi) from 1 to 4.

The experiment measured the power consumption, message delay, and additional message length (bytes). Moreover, the analysis was also used to model security algorithms. The experimental results from long runs (72 hours) reveal the cryptographic solution choice is significant for the message delay and additional message length. Additionally, comparative analyses of diverse symmetric block-based cryptographic algorithms have been conducted in [8] to aid in algorithm selection tailored to specific IoT applications. Evaluations encompassing AES, CLEFIA, DES, Triple DES, IDEA, PRESENT, SEA, SPECK, TEA, XTEA, and TWOFISH shed light on energy, power, and memory consumption, alongside throughput considerations. Also, by assessing efficiency and security parameters, the comparative analysis in [2] seeks to identify the optimal cryptographic algorithm, namely between AES, DES, RSA and lightweight cryptographic algorithm Fernet, capable of minimizing privacy and security risks in IoT applications, thereby safeguarding the integrity of IoT data amidst evolving threat landscapes. Moreover, the feasibility of deploying lightweight cryptography algorithms in resource-constrained smart devices has been explored, acknowledging the challenges posed by the need for cheap, low-power IoT nodes with reduced computational performance. Through experimental performance in [7] analyses of several cryptography solutions targeted for embedded platforms, researchers aim to assist IoT developers in selecting suitable algorithms based on efficiency metrics such as execution time, code efficiency, and communication influences. The results of this work intend to guide IoT developers in choosing appropriate cryptography algorithms tailored to the specific requirements of their IoT applications.

Table I presents a comparison between the works cited and the model proposed in this work. For the variation size, this column indicates if the article presented data on various packets sizes, for example MB, KB and others. The variation of algorithms column indicates if the article presented more than one algorithm. Relatively to the times column, it indicates if the article uses encryption and decryption time to compare different algorithms. Regarding the integration column, it indicates if the article used cloud computing in their computing in their tests. Finally, both the CPU and Memory columns indicate if the article mentioned memory usage and information about the impact of different types of CPU in the cryptographic solution.

In addition to the related works compared and presented in table 1, other scenarios and applications are based on the works developed in [19-24]. These works present scenarios and discussions as well as descriptions, criteria, parameters and concrete definitions of investigations carried out on the topic and area, which require added value related to encryption and data management.

TABLE I - COMPARTIVE ANALYSIS

| Work | Variation in size | Variation of algorithms | Times | Integration | CPU | Memory |
|---|---|---|---|---|---|---|
| [1] | | X | | | | |
| [2] | | X | X | | | |
| [4] | | X | | | | |
| [7] | | X | X | | | |
| [8] | | X | X | | | X |
| [13] | | X | X | | X | X |
| [14] | | X | X | | | |
| This work | X | X | X | X | X | X |

Building upon the insights gained from these studies, our study aims to also create a robust cryptographic model, but different from the models created in the articles, our structure model will use encapsulation to provide a better data security. The encapsulation process will be realized with different cryptographic algorithms to not allow attackers to easily identify the pattern used, so if a plain text is encrypted with AES then the result will be encrypted with another algorithm. This study will also be integrating cloud computing to confront the challenges posed by the limitations of computational resources in Iot devices.

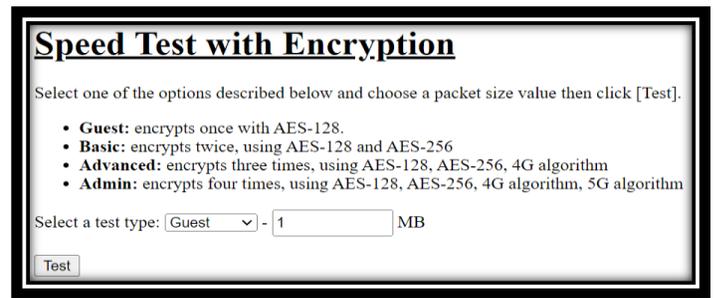

Fig. 1. First prototype and interface created.

## III. Methodology

The objective of this study was to enhance the security and performance of data transmission in interconnected IoT devices. This was achieved by analyzing, selecting, and implementing efficient cryptographic algorithms and leveraging cloud computing to optimize processing efficiency and resource utilization. The cryptographic algorithms selected for this study were chosen based on a comprehensive literature review to identify current cryptographic algorithms used in IoT applications, especially those applicable in 5G network scenarios it was also integrated a 4G network algorithms to link to other studies that were made in scenarios and applications that uses 4G network. The selected algorithms were evaluated based on their performance on the structure model in terms of encryption and decryption times, packet size handling, packet quantity and overall performance. The primary algorithms selected for this study included AES [9] (Advanced Encryption Standard) being the modes used CBC and CTR, Blowfish [9], Chacha20 and XChacha20 [10], ECDH[11] (Elliptic Curve Cryptography Diffie–Hellman), and HMAC-SHA-256 [3] (Hash-based Message Authentication Code using SHA-256).

The implementation phase involved creating a program to test different cryptographic algorithms in terms of performance. Developed in Python, using the PyCryptodome library, a multi-layered encryption model was created to use encapsulation with the chosen cryptographic algorithms. This structure model is designed with four security levels: Guest, Basic, Advanced, and Admin, each corresponding to an increasing degree of security, with higher levels involving more complex encryption processes. For instance, data encrypted at the Advanced level undergoes triple encryption, enhancing security at the cost of increased computational requirements and longer processing times.

Structured models were tested on two machines with varying specifications to determine the best performance. Initial testing involved creating various structured models using the selected algorithms to find the optimal configuration for encryption and decryption times, packet quantity, and size. Tests were performed on machines with AMD Ryzen processors and varying RAM capacities.

In the initial testing phase, the encryption and decryption times for different packet sizes and algorithms combinations were evaluated. This phase also assessed the effect of varying packet quantities on encryption and decryption times. The structured model was iteratively refined based on these tests to identify the optimal combination of algorithms for both encryption and decryption.

The next phase was the integration of cloud computing that played a crucial role in mitigating the computational limitations of IoT devices by offloading intensive cryptographic operations, by allowing the separation of both processes (encryption and decryption), to the cloud. The performance of structure model in a cloud environment was analyzed using Microsoft Azure and Raspberry Pi as the primary tools for implementation. This analysis focused on measuring key metrics such as encryption and decryption times, impact of packet size, and packet quantity, memory used in each level, client and server process time. This approach aimed to optimize processing efficiency and resource utilization, allowing for scalable and flexible implementation of the cryptographic model. The system was designed with a virtual machine in the cloud acting as the host for the client program responsible for the decryption process, while a Raspberry Pi handled the encryption process and communicated with the cloud-based virtual machine, which is illustrated in the figure 2. It was also created a new interface for this cryptosystem as it is shown in figure 3.

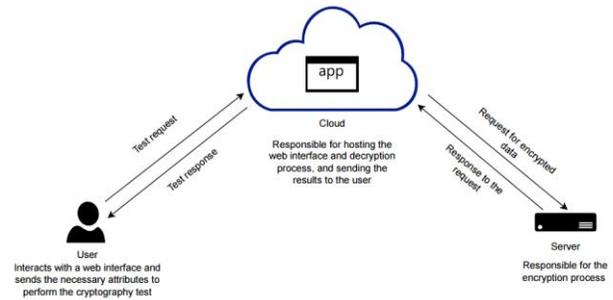

Fig. 2. Cryptosystem created.

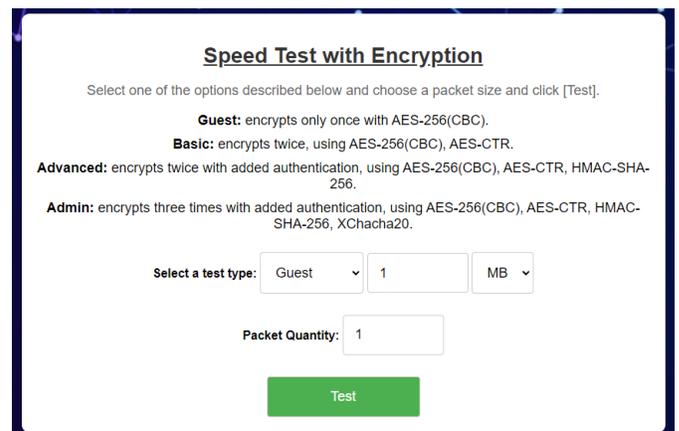

Fig. 3. New interface implemented.

Through analysis of previously stages, it was possible to conclude that since a machine with 32 GB was responsible for both the encryption and decryption of packets at the begging, then in a cryptosystem where one machine does the encryption while the other does the decryption, it was determined that both machines should have at least 16 GB RAM to achieve the packet quantity that was observed when using only one machine. However, due to some limitations it was only possible to create one virtual machine in Microsoft Azure with 8 GB RAM

## IV. Results

The initial testing phase focused on evaluating the encryption and decryption times for different packet sizes using the initial structure model, fig.1, with Chacha20(4G) and Blowfish(5G). The admin and Guest level were used for drawing conclusions since they represented the simplest and most complex levels respectively. The packet sizes tested included 1, 5, 10, 20, and 50 MB. The results showed that both encryption and decryption times increased with packet size. Specifically, the encryption and decryption times for larger packet sizes were significantly higher

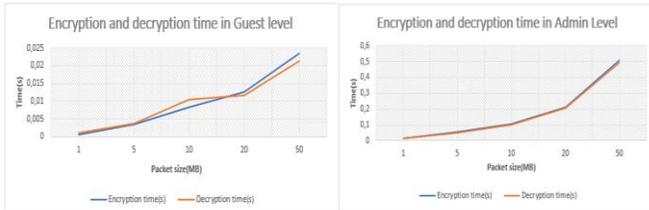

Fig. 4. Performance of the initial structure model.

Overall, it is clear that the encryption and decryption times are substantially lower at the Guest level compared to the Admin level. The Guest level shows the best performance with the smallest packet sizes (1 MB) having the quickest processing times. On the other hand, the Admin level, although having longer times, reflects the increased security required for higher levels of access, in which the packet size 1 MB also had the quickest processing.

The next phase involved substituting Blowfish or ChaCha20 with AES-CTR in the structure model. This modification aimed to enhance performance while maintaining security. Chacha20 together with AES-CTR showed initially, for smaller packets, showed similar performances to the combination of Blowfish and AES-CTR. The difference became more pronounced with larger packet sizes, with Chacha20 taking 265 milliseconds for a 50 MB packet compared to Blowfish's 397 milliseconds. Similarly, Chacha20 outperformed Blowfish in decryption times. For a 1 MB packet, Chacha20's decryption time was around 5 milliseconds, while Blowfish's was 7 milliseconds. For a 50 MB packet, Chacha20 took 290 milliseconds, and Blowfish took 407 milliseconds.

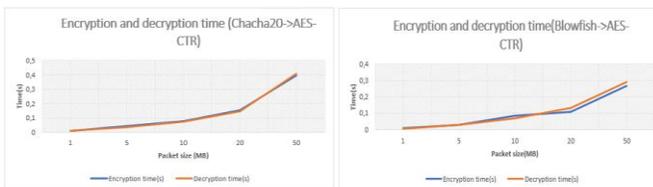

Fig. 5. Performance of Chacha20 and Blowfish with AES-CTR.

A new structure model was created using AES-CTR, along with the introduction of ECDH and HMAC-SHA-256 at the admin level, these were introduced one at time. The analysis of the results indicated that while ECDH provided good security, it increased the processing time when combined with AES-CTR. The introduction of HMAC-SHA-256 at the beginning of the process also impacted the performance.

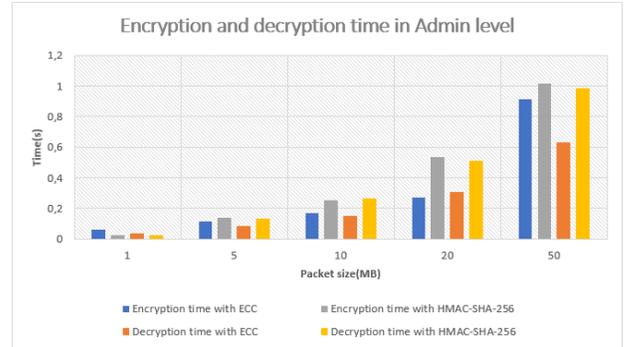

Fig 5. ECDH VS HMAC-SHA-256

For smaller packet sizes (1 MB and 5 MB), HMAC-SHA-256 shows better performance in the structure model than ECDH. However, as the packet size increases, ECDH demonstrates a consistently better performance. At 10 MB, ECDH's encryption time is 0.16 seconds compared to when the structure mode used HMAC-SHA-256 for authentication takes 0.25 seconds. For 20 MB packets, ECDH takes 0.27 seconds while the structure model with HMAC-SHA-256 takes 0.53 seconds. The largest packet size (50 MB) highlights the most significant difference, with ECDH at 0.9 seconds and HMAC-SHA-256 at 1.01 seconds. The decryption time of the structure model also favors ECDH, especially as the packet size increases as it can be seen in the graph in Fig.5.

Finally, to identify the optimal structure model after all the tests that were previously done, evaluations were made to find which algorithms would be better for each level, since as demonstrated before just because a algorithm worked with multiple algorithms doesn´t mean it's the best, for example for the Guest level that uses only one algorithm so it was necessary to analyze the time taken for encryption and decryption of different packets sizes with different combinations of algorithms. Below is presented two of the graphs that were created with the results of the tests realized to find the best structure model, it is only presented the encryption and decryption time in the Guest level for a better interpretation of what was done.

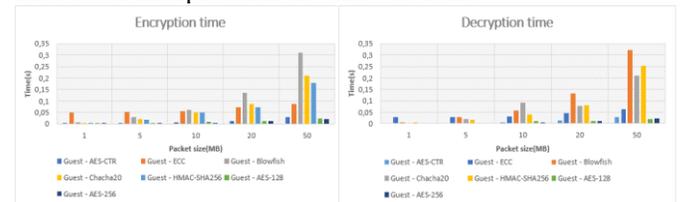

Fig 5. Performance of different algorithms in the Guest Level

It´s possible to observe that AES-256 consistently shows the best performance across all packet sizes. For instance, at 1 MB, AES-256 encryption time is approximately 0.0013 seconds, significantly lower than the other algorithms. As packet sizes increase, AES-256 maintains its efficiency. At 50 MB, AES-256 encryption time is approximately 0.0219 seconds, whereas other algorithms such as ECC take approximately 0.08 seconds and Blowfish takes 0.31 seconds.

For the next phase, the impact of varying packet quantities on encryption performance was evaluated. The structure model that was used in this phase was the one that had the best performance when it came to encryption and decryption time. In this phase the encryption and decryption process were also realized in the same machine, one with 16GB RAM and the other with 32 GB RAM.

Firstly, it was tested the packet quantity on the machine with 16 GB of RAM, and it was determined that, to find the limitation of the packet quantity associated with computational limitations, the level Admin was used as reference since is the most complex of the four levels. Below it´s possible to observe the graphs that correspond to the encryption time and decryption time in the admin level with packet quantity, for this test the packet sizes that were used were 1,5,10,20,50 MB.

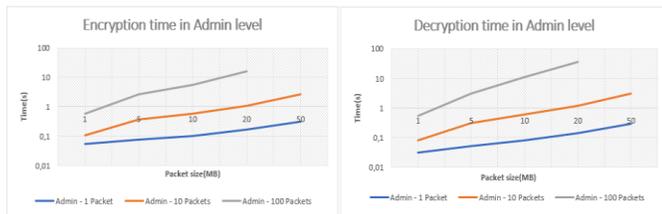

Fig 6. Impact of packet quantity

It can be observed in the encryption graph on fig.6 the encryption times for 1, 10, and 100 packets as the packet size increases, the encryption time increases almost linearly. The encryption time for 1 packet remains the lowest, while 100 packets exhibit the highest times across all packet sizes. It is observed similar trends for decryption times, with decryption taking longer as the packet size increases. The decryption graph also highlights the differences in decryption times between handling 1, 10, and 100 packets. So, in summary, the more packets are used more time it takes to encrypt and decrypt them which could compromise the efficiency of our structure model. Through analysis of both graphs, it can be observed that with both process on the same machine and 16 GB of RAM, it is possible to transmit 10 packets of 50 MB and 100 packets of 20 MB, this is the first limitation when utilizing both process on the same machine and with 16 GB of RAM.

Secondly, it was tested the same structured model but on a machine with 32 GB of RAM and the results can be seen the graphs below.

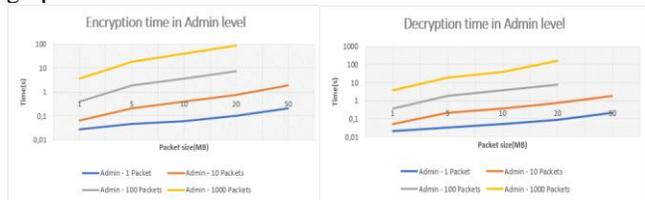

Fig 7. Impact of packet quantity on a new machine

On fig.7 the graphs for, the encryption and decryption times for 1, 10, 100, and 1000 packets are depicted. It is noticeable that there is a reduction in encryption and decryption time with the increased RAM for the transmission of 1,10,100 packets, although the general trend of increasing time with packet size remains. Through observation of both graphs, it is possible to deduce a new limitation of the model which is 10 packets of 50 MB and 1000 packets of 20 MB.

Finally, it was decided to introduce a new packet size to compensate for the limitations when dealing with packets with values of MB and packet quantity. The new packet size introduced was KB, and since it a smaller packet it was possible to transmit more packets at the same time, the values used for the tests were 1,5,10,15 to 120 KB and the tests were realized in the machine with 32 GB of RAM. which is represented in the graphs below for the Admin level.

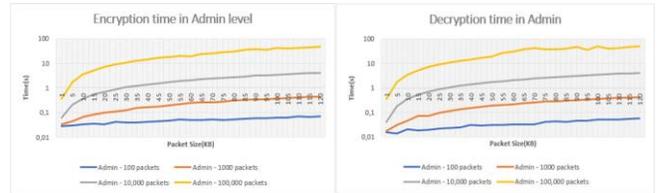

Fig 8. Impact of packet quantity on KB packets

Through visualization of both graphs in Fig 8., it´s possible to see that with smaller packets it´s possible to transmit more packets at the same time. However, in these tests both the encryption process and decryption process were realized on the same machine which means that the time taken to encrypt, and decrypt will be bigger since the resources of the same machine are being used for two processes at the same time. The provided figures collectively emphasize the impact of packet size and quantity on the encryption and decryption performance. Larger packet sizes and higher quantities result in increased processing times. Machines with higher RAM capacities significantly mitigate these times, demonstrating the importance of hardware resources in cryptographic computations for IoT applications.

Last phase of this study was to analyze the cryptosystem created, However, due to some limitations it was only possible to create one virtual machine in Microsoft Azure with 8 GB RAM, with that said the application that the user interacts with and is responsible for the decryption process is hosted in the cloud while the raspberry pi, that acts as a server, is responsible for encryption process, besides that it was also necessary to replace ECDH due to certain circumstances being the new algorithm XChacha20.

The client-side application, built with Flask, includes an endpoint for sending test data to the server and another for decrypting the data received and displaying the results.

```python
@app.route('/test_speed', methods=['GET', 'POST'])
def test_speed():
    test_type = request.form['test_type']
    custom_size = int(request.form['custom_size'])
    value_unit = request.form['unit']
    packet_quantity = int(request.form['packet_quantity'])

    response_data = send_data(test_type, custom_size, value_unit, packet_quantity)
    clientTimeProcess_start = time.time()
    encrypted_data = base64.b64decode(response_data['encrypted_data'])

    key_map = {
        'Guest': ['key'],
        'Basic': ['key_aes256', 'key_aesCtr'],
        'Advanced': ['key_aes256', 'key_aesCtr', 'key_hmacSha256'],
        'Admin': ['key_aes256', 'key_aesCtr', 'key_hmacSha256', 'key_xchacha20', 'nonce']
    }

    keys = {key: base64.b64decode(response_data[key]) for key in key_map[test_type]}

    decryption_start_time = time.time()

    if test_type == 'Guest':
        decrypted_data = ed.aes_256_decrypt(encrypted_data, keys['key'])
    elif test_type == 'Basic':
        decrypted_data = ed.aes_256_decrypt(
            ed.aes_ctr_decrypt(encrypted_data, keys['key_aesCtr']),
            keys['key_aes256']
        )
```

Fig 9. Client process code

In fig.9 it is possible to observe how the code was created, it is only presented the decryption process of the Guest and Basic level but the code for the other levels follows the same logic. This program handles the HTML form submission from the client, sends data to the server, and processes the response. The server-side application, also built with Flask, includes an endpoint to receive data from the client, process it, and send back the results to the client. The metrics studied on this final phase were the time of encryption and decryption, client and server process time and lastly memory and CPU used.

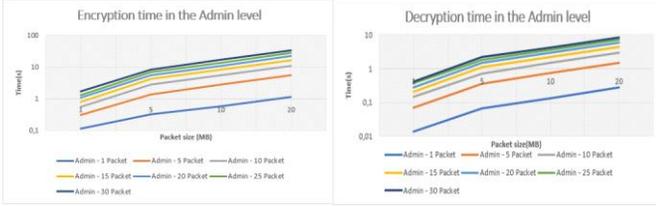

Fig. 10. Performance of encryption ad decryption in the cryptosystem

The graphs, in fig.10, illustrate the encryption and decryption time for various packet quantities (1, 5, 10, 15, 20, 25, and 30 packets). The results indicate that the encryption and decryption time increases with both the packet size and the number of packets. The lines for each packet quantity show a consistent upward trend, reflecting the increased computational load as the data size grows.

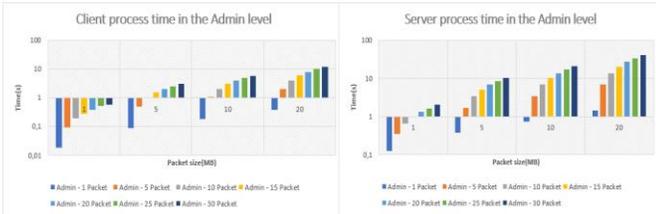

Fig. 11. Performance of encryption ad decryption in the cryptosystem

The graphs in fig.11, it illustrates the time taken for the client and server to finish all operations. The data shows that client processing time increases with both the packet size and the number of packets. The chart highlights significant differences in processing times as packet size grows, with more pronounced increases for higher packet counts. The server, however, takes longer to process larger packet sizes and higher quantities of packets. The bars representing each packet quantity demonstrate a clear escalation in processing time as the data load increases.

With the four graphs presented it´s possible to observe that when it comes to both processes the client has a better performance than the server, since both have 8 GB RAM the reason for this difference lies in the difference of CPU used since the virtual machine in the cloud has an Intel processor while the raspberry pi has ARM processor, which as demonstrated impacts the time of encryption, decryption, client process and server process.

Besides these tests, the memory used for both sides, client-server, was also tested in the four levels of the structure model for 20 MB and 10,20,30 packets, the results can be observed below for both sides.

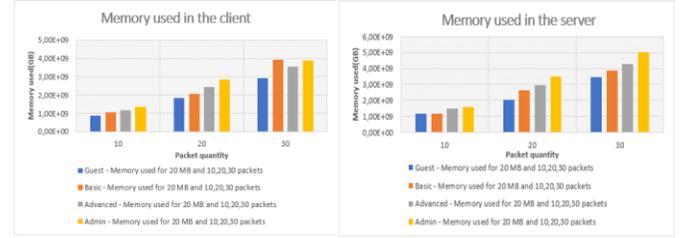

Fig. 12. Memory usage in the cryptosystem

As it is possible to observe in fig.12, the memory consumption as the packet quantity rises from 10 to 30. Among the access levels, the Guest level consistently shows the lowest memory usage for all packet quantities. The Basic level exhibits a moderate increase in memory consumption compared to the Guest level, while the Advanced level shows a further rise, indicating more complex processing requirements. The Admin level, being the most resource-intensive, displays the highest memory usage across all packet quantities Overall, these graphs highlight that both packet quantity and the level used significantly impact memory usage. The server requires more memory than the client, especially at the Admin level and with higher packet quantities. This analysis is crucial for understanding the resource requirements and optimizing the performance of cryptographic operations in our test scenario.

## V. CONCLUSION

In this study, we have addressed the significant concern of ensuring data security in interconnected IoT devices, particularly within 5G networks. By analyzing and selecting efficient cryptographic algorithms, we designed a model that optimizes data transmission while maintaining robust security. The integration of cloud computing played a crucial role in enhancing processing efficiency and resource utilization, ultimately leading to better data security and transmission performance. The scientific contribution is a hierarchical security model that features four levels: Guest, Basic, Advanced, and Admin. Each level, except Guest, uses encapsulation with specific cryptographic algorithms, tailored to the sensitivity of data and user privileges. By leveraging cloud computing, we enhanced the processing capabilities and resource utilization of the structured model. This integration allowed us to offload complex cryptographic computations to the cloud, thereby reducing the burden on individual IoT devices. Finally, through our research, we optimized cryptographic processes to reduce computational overhead without compromising security. This optimization is crucial for the practical deployment of secure IoT networks. Regarding the results, it was determined that the best structured model used the algorithms AES-256, AES-CTR, HMAC-SHA-256 and XChacha20 since these four algorithms had a good performance throughout the tests made, besides this study managed to raise the amount of data transmitted relatively to another study done in the same area. With this structured model it was possible to transmit 1000 packets of 20 MB, it is important to note that this achievement was possible on a single machine of 32 GB doing both encryption and decryption, by dividing the process that same achievement should be possible, but it is necessary 16 GB on both sides, so it's important to explore further the memory management since all comes with costs. All the resusts obtained in this study, as well as all the graphs and times can be found in [18].

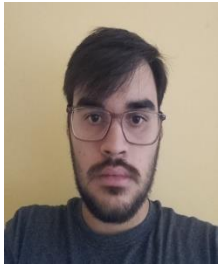

**P.M.C.F da costa** is currently in the last year of the degree of Electronic, Telecommunications, and Computers Engineering at Instituto Superior de Engenharia de Lisboa (ISEL). With a keen interest in programming and cybersecurity, he is eager to deepen his knowledge and skills in these areas, preparing for a successful career in the ever-evolving field of technology.

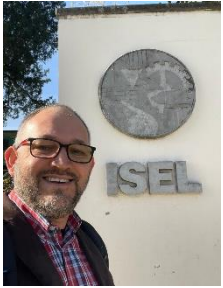

**V. R. Q. Leithardt** is Profesor the Instituto Superior de Engenharia de Lisboa (ISEL), Lisboa – Portugal, and Senior Member of IEEE. He received the Ph.D. degree in Computer Science from INF-UFRGS, Brazil, in 2015. He is also an Integrated Member of the Center of Technology and Systems (UNINOVA-CTS) and Associated Lab of Intelligent Systems (LASI), 2829-516 Caparica, Portugal.